\documentstyle[preprint,aps,epsfig]{revtex}

\newcommand{\be}{\begin{equation}}
\newcommand{\ee}{\end{equation}}
\newcommand{\Dlt}{\Delta}
\newcommand{\dlt}{\delta}
\newcommand{\om}{\omega}

\newcommand{\br}{{\bf r}}
\newcommand{\vp}{\varphi}
\newcommand{\ep}{\varepsilon}

\newcommand{\al}{\alpha}

\newcommand{\ra}{\rightarrow}

\newcommand{\prt}{\partial}

\tightenlines

\begin{document}

\draft

\title{Entanglement production with multimode Bose-Einstein condensates 
in optical lattices} 
\author{V.I. Yukalov$^{1,2}$ and E.P. Yukalova$^{1,3}$} 

\address{$^1$Institut f\"ur Theoretische Physik, \\
Freie Universit\"at Berlin, Arnimallee 14, D-14195 Berlin, Germany}
\address{$^2$Bogolubov Laboratory of Theoretical Physics, \\
Joint Institute for Nuclear Research, Dubna 141980, Russia}
\address{$^3$Department of Computational Physics, Laboratory of Information
Technologies, \\
Joint Institute for Nuclear Research, Dubna 141980, Russia}

\maketitle

\vskip 2cm

\begin{abstract}

Deep optical lattices are considered, in each site of which there are 
many Bose-condensed atoms. By the resonant modulation of trapping potentials 
it is possible to transfer a macroscopic portion of atoms to the collective 
nonlinear states corresponding to topological coherent modes. Entanglement 
can be generated between these modes. By varying the resonant modulating
field it is possible to effectively regulate entanglement production in 
this multimode multitrap system of Bose condensates.

\end{abstract}

\newpage

Entanglement is assumed to play an important role in quantum computation 
and quantum information processing [1--5]. As possible candidates for 
engineering entanglement, one considers spin assemblies, trapped ions, 
or Bose-condensed trapped atoms. Here we shall concentrate our attention 
on atomic Bose-Einstein condensates [6--9]. We shall study entanglement 
production that can be realized between different topological coherent 
modes of nonequilibrium Bose condensates. A general theory of the resonant 
generation of these coherent modes was advanced in Ref. [10]. The properties
of these modes have been intensively investigated [10--30]. A brief survey 
is given in a recent publication [31].

Let us, first, recall the meaning of the topological coherent modes [10]. 
We consider a dilute Bose gas, with the effective local interaction
\be
\label{1}
\Phi(\br) = \Phi_0\dlt(\br) \; , \qquad 
\Phi_0 \equiv 4\pi\; \frac{a_s}{m} \; ,
\ee
where $a_s$ is a scattering length, $m$ is atomic mass. When an atomic 
system is confined in a trap and cooled down to very low temperatures,
then almost all atoms can be piled down to a single quantum state, thus 
creating Bose-Einstein condensate. The latter forms a coherent system, 
which is described by the Gross-Pitaevskii equation
\be
\label{2}
\hat H[\eta_k]\; \eta_k(\br) = E_k\;\eta_k(\br) \; ,
\ee
with the nonlinear Schr\"odinger Hamiltonian
\be
\label{3}
\hat H[\eta] \equiv -\; \frac{\nabla^2}{2m} + U(\br) + \Phi_0|\eta|^2 \; ,
\ee
in which $U(\br)$ is a trapping potential. The solutions to the eigenproblem 
(2), labelled by a set of quantum numbers $k$, are the topological coherent 
modes $\eta_k(\br)$. These are not compulsorily orthogonal, but can always 
be normalized to the total number of atoms in the trap,
$$
||\eta_k||^2 = (\eta_k,\eta_k) = N \; .
$$
The lowest-energy mode $\eta_0(\br)$ represents the usual Bose-Einstein 
condensate, while the higher-energy modes describe nonground-state 
condensates.

For each coherent mode $\eta_k(\br)$, one can construct the coherent 
state
\be
\label{4}
|\eta_k> \; = \left [ \frac{e^{-N/2}}{\sqrt{n!}} \; \prod_{i=1}^n
\eta_k(\br_i) \right ] \; ,
\ee
which is a column with elements labelled by the index $n$. The coherent 
states are normalized to unity, $<\eta_k|\eta_k>=1$. Though, in general, 
coherent states are not orthogonal, it is possible to show [32] that the 
states (4) are asymptotically orthogonal, in the sense that
$$
\lim_{N\ra\infty}\; <\eta_k|\eta_p> \; = \; \dlt_{kp} \; .
$$
The totality of all $|\eta_k>$, that is, the closed linear envelope 
$\overline{\cal L}\{|\eta_k>\}$, forms the space of states for the trapped 
atomic system.

Now, let us consider a set of traps, in each of which there are many 
Bose-condensed atoms. Such a setup can be realized by creating a deep 
optical lattice. For example, cold rubidium $^{87}$Rb atoms have been 
loaded [33] into an optical lattice, with adjacent sites spaced by 
$a=5.3\times 10^{-4}$ cm. Lattice sites were practically independent, 
with the tunneling time between sites above $10^{18}$ s. The total number 
of Bose-condensed atoms in the optical lattice was about 7000. The number 
of lattice sites was typically between 5 to 35. So that the number of 
condensed atoms in each site could be varied between about 200 to 2000.

Nowadays there exists a variety of optical lattices, one-dimensional, 
two-dimensional, or three-dimensional [34,35]. Multiphoton processes [36]
can be employed for creating asymmetric lattice potentials [37]. Optical
lattice potentials can be made spin-dependent [38]. Being separated in 
different lattice sites, atoms are practically independent. But if there 
occurs small, though finite, tunneling, phase coherence may persist on 
short length scales even deep in the insulating state [39].

Assume that in each lattice site of an optical lattice there are many 
Bose-condensed atoms, so that each site plays the role of a trap. The 
space of states for a $j$-trap is
\be
\label{5}
{\cal H}_j = \overline{\cal L}\{|\eta_{jk}>\} \; .
\ee
The total space of states for the whole lattice is
\be
\label{6}
{\cal H} = \otimes_j {\cal H}_j \; .
\ee
The set of disentangled states,
\be
\label{7}
{\cal D} \equiv \{ \otimes_j|\vp_j> : \; |\vp_j> \in{\cal H}_j \} \; ,
\ee
consists of the states having the form of the tensor products. An arbitrary 
state of the considered system of traps can be represented as
\be
\label{8}
|\eta(t)> \; = \sum_k \; c_k(t) \otimes_j|\eta_{jk}> \; ,
\ee
whose coefficients define the mode probabilities, or the fractional mode 
populations
\be
\label{9}
n_k(t) \equiv |c_k(t)|^2 \; .
\ee

Let the number of trapping sites be $L$, with $M$ coherent modes each. 
If the coefficients $c_k(t)$ in state (8) can be varied, then different 
entangled states can be created, such as the Bell states
$$
|B>\; = \frac{1}{\sqrt{2}} (|00> \pm \; |11>)
$$
for $L=2$ and $M=2$, the multicat states
$$
|MC>\; = \frac{1}{\sqrt{2}} ( |00\ldots 0> \pm \; |11\ldots 1>)
$$
for $L>2$ and $M=2$, or the multicat multimode states for $L>2$ and 
$M>2$.

When the system is in a pure statistical state, then its statistical 
operator is
\be
\label{10}
\hat\rho(t) = |\eta(t)><\eta(t)| \; .
\ee
In the case of a mixed statistical state, the statistical operator 
becomes
\be
\label{11}
\hat\rho(t) = \sum_k |c_k(t)|^2 \otimes_j |\eta_{jk}><\eta_{jk}| \; .
\ee

Entanglement, generated by a statistical operator $\hat\rho(t)$, is 
quantified [40,41] by the measure of entanglement production
\be
\label{12}
\ep\left ( \hat\rho(t)\right )  \equiv \log\; 
\frac{||\hat\rho(t)||_{{\cal D}}}{||\hat\rho^\otimes(t)||_{{\cal D}}}\; ,
\ee 
in which the logarithm is to the base 2,
$$
||\hat\rho(t)||_{{\cal D}} \equiv \sup_{f\in {\cal D}} ||\hat\rho(t)f||
\qquad (||f||=1) \; ,
$$
$$
\hat\rho^\otimes(t) \equiv \otimes_j \hat\rho_j(t)\; , \qquad
\hat\rho_j(t) \equiv {\rm Tr}_{{\cal H}_{i\neq j}}\hat\rho(t) \; .
$$
For $L$ trapping sites of a lattice, we find
\be
\label{13}
\ep\left (\hat\rho(t)\right ) = (L-1)\ep_2(t) \; ,
\ee
where
\be
\label{14}
\ep_2(t) = -\log\; \sup_k \; n_k(t) \; .
\ee

In order to generate higher coherent modes, it is necessary to apply an 
external resonant field [10,31]. We assume that the same modulating field 
acts on all lattice sites. The field has the form
\be
\label{15}
V(\br,t) = \zeta(t) \left [ V_1(\br)\cos(\om t) + V_2(\br)\sin(\om t)
\right ] \; .
\ee
Here $\zeta(t)=0,1$ is a switching function that allows one to switch on 
and off the resonant field (15). The frequency $\om$ is tuned close to the 
transition frequency
\be
\label{16}
\om_0 \equiv E_1 - E_0 \; ,
\ee
in which $E_0$ is the energy of the ground-state mode and $E_1$ is the 
energy of the desired mode to be generated. The resonance condition is 
implied, such that
\be
\label{17}
\left | \frac{\Dlt\om}{\om} \right | \ll 1 \qquad 
(\Dlt \equiv \om-\om_0) \; .
\ee
With the resonant modulating field (15), the time-dependent Gross-Pitaevskii
equation becomes
\be
\label{18}
i \; \frac{\prt}{\prt t} \; \vp(\br,t) = \left ( \hat H[\vp] +
\hat V\right )\vp(\br,t) \; ,
\ee
where the nonlinear Schr\"odinger Hamiltonian is
$$
\hat H[\vp] = -\; \frac{\nabla^2}{2m} + U(\br) + \Phi_0 N|\vp|^2
$$
and the function $\vp(\br,t)$ is normalized to one, $||\vp||^2=1$.

The solution to Eq. (18) can be represented as the mode expansion
\be
\label{19}
\vp(\br,t) = \sum_k c_k(t) \vp_k(\br) e^{-i E_k t} \; ,
\ee
in which the coefficient functions $c_k(t)$ define the fractional mode 
populations (9). These functions can be found as the projections
$$
c_k(t) = \frac{1}{T} \; \int_t^{t+T} \; \vp_k^*(\br) e^{iE_kt}
\vp(\br,t)\; dt \qquad \left ( T =\frac{2\pi}{\om} \right ) \; .
$$

The dynamics of the mode populations essentially depends on the strength 
of atomic interactions and on the amplitude of the resonant field [10,31]. 
We shall denote by $b$ the dimensionless amplitude of the modulating resonant
field, reduced to the strength of atomic interactions. Respectively, the
temporal behaviour of the entanglement production measure (14), which for 
the two-mode case takes the form
\be
\label{20}
\ep_2(t) = -\log_2\; \sup\{ n_0(t),\; n_1(t)\} \; ,
\ee
strongly depends on the properties of the resonant field.

We assume that at the initial time $t=0$ solely the ground-state is 
available, so that $n_0(0)=1$ and $n_1(0)=0$. Then the resonant field 
(15) is switched on, generating an excited coherent mode and changing the
entanglement production measure (20). There are two ways of regulating 
the amount of entanglement production.

First, one can vary the amplitude and frequency of the resonant pumping 
field, choosing by this the required parameters, whose values define two 
main regimes of an oscillatory behaviour of $\ep_2(t)$. These are the 
mode-locked and mode-unlocked regimes [31]. We solve numerically the
evolution equations and calculate the measure of entanglement production 
(20). Keeping in mind that the detuning can always be made small, we set 
it to zero. Then the value $b_c=0.497764$ is the critical point for the 
change of the dynamical regimes. Below $b_c$, the measure (20) oscillates 
with time, never reaching one, as is shown in Fig. 1. When the parameter 
$b$ reaches the critical point $b=b_c$, then the oscillating $\ep_2(t)$ 
reaches one, as is demonstrated in Fig. 2. For the dimensionless amplitude 
of the pumping field $b>b_c$, the oscillations of $\ep_2(t)$ are always in 
the interval between 0 and 1. But the oscillation period sensitively depends 
on the value of $b$. Thus, for $b=0.5$, just a little above $b_c$, the 
period of oscillations, shown in Fig. 3, is more than twice shorter than 
that in Fig. 2 for $b=b_c$. The period for $b=0.7$, as is demonstrated in 
Fig. 4, is about eight times shorter than in Fig. 2. Thus, by varying the 
amplitude of the pumping resonant field, we can strongly influence the 
evolution of $\ep_2(t)$ both in its amplitude and period of oscillations.

There is one more very interesting way of regulating entanglement 
generation, which can be done by switching on and off the applied 
resonant field. Recall that this alternating field can be easily produced 
by modulating the magnetic field forming the trapping potential in magnetic 
traps or by varying the laser intensity in optical traps. Then, it is 
possible to create various sequences of pulses for $\ep_2(t)$. For example, 
by switching on and off the resonant field in a periodic manner, we may 
form equidistant pulses of $\ep_2(t)$, with all pulses having the same shape, 
as is demonstrated in Fig. 5. But we can also switch on and off the pumping 
field at different time intervals, thus, forming nonequidistant pulses, as 
is shown in Fig. 6. The possibility of creating very different pulses is 
illustrated in Fig. 7. Regulating entanglement production by means of a 
manipulation with the resonant pumping field, it is feasible to organize 
a kind of the Morse alphabet.

Here we have considered entanglement production in a multimode Bose-Einstein 
condensate loaded in a deep optical lattice. The arising entanglement 
production occurs for different coherent modes generated by a resonant 
external field. Recently ultra-cold fermions in optical lattices have 
attracted great attention (see survey [42]). Entanglement production 
for fermions in optical lattices can also be considered, though requiring 
different techniques. Varying the interaction between fermions by means 
of the Feshbach resonance methods, bound fermionic states can be achieved, 
forming bosonic molecules. The latter can be Bose-condensed (see [43,44] 
and references therein). Therefore the coherent modes of molecules can be 
created, similarly to those of atoms. Then the entanglement production for 
molecular coherent modes can be studied in the way analogous to that for 
atomic coherent modes.

The measure of entanglement production (14) or (20) is directly connected 
with the fractional mode populations. The latter define the spatial features 
of atomic clouds inside each lattice site. Thus, analyzing the spatial 
distribution of atoms, one can make conclusions on the mode entanglement 
production. The spatial distribution can be studied by means of scattering 
experiments. Wave scattering on periodic structures is known to possess a
number of interesting properties [45]. Another possibility of studying the 
spatial characteristics of atomic clouds is through the time-of-flight
experiments, by releasing atoms from the trapping potentials and observing
the atom expansion and interference.

\vskip 2mm

In conclusion, a multitrap ensemble of multimode Bose-Einstein condensates, 
subject to the action of a common resonant field, is analogous to a system 
of finite-level atoms in a common resonant electromagnetic field. A multitrap 
system can be formed, e.g., as an optical lattice with deep potential wells, 
incorporating many atoms around each lattice site. In the multitrap multimode 
condensate a high level of entanglement can be achieved. By varying the 
amplitude and frequency of the pumping resonant field, different regimes 
of evolutional entanglement can be realized. Moreover, by switching on and 
off the pumping field in various ways, it is feasible to create entanglement 
pulses of arbitrary length and composing arbitrary sequences of {\it 
punctuated entanglement generation}. Such a high level of admissible 
manipulation with and regulating of entanglement can, probably, be useful 
for information processing and quantum computing.

\vskip 3mm

{\bf Acknowledgement}

\vskip 2mm

Financial support form the German Research Foundation is appreciated. 
One of us (V.I.Y.) is grateful to the German Research Foundation for the 
Mercator Professorship.

\newpage

\begin{center}
{\large{\bf Figure Captions}}
\end{center}

\vskip 2cm

{\bf Fig. 1}. Evolutional entanglement production, quantified by the 
measure $\ep_2(t)$, in the mode-locked regime, with $b=0.3$. In this and 
in all following figures, time is measured in units of $\al^{-1}$.

\vskip 1cm

{\bf Fig. 2}. The measure $\ep_2(t)$ for the boundary between the 
mode-locked and mode-unlocked regimes, when $b=b_c=0.497764$.

\vskip 1cm

{\bf Fig. 3}. Entanglement production in the mod-unlocked regime, with 
$b=0.5$.

\vskip 1cm

{\bf Fig. 4}. Drastic shortenning of the period of $\ep_2(t)$ for 
$b=0.7$.

\vskip 1cm

{\bf Fig. 5}. Regulated equidistant pulses of $\ep_2(t)$, formed by 
switching on and off the resonant field, with $b=0.7$, so that $\ep_2(t)$ 
equals one during the time intervals $\Dlt t=7.35$ (in units of $\al^{-1}$),
and it equals zero during the same intervals $\Dlt t=7.35$.

\vskip 1cm

{\bf Fig. 6}. Nonequidistant pulses of $\ep_2(t)$, created by switching 
on and off the pumping field, with $b=0.7$, at nonequal time intervals.

\vskip 1cm

{\bf Fig. 7}. Regulated pulses of $\ep_2(t)$, for the same $b=0.7$, as 
in Fig. 6, but for essentially different moments of switching on and off
the pumping field.

\newpage

\begin{figure}[ht]
\centerline{\psfig{file=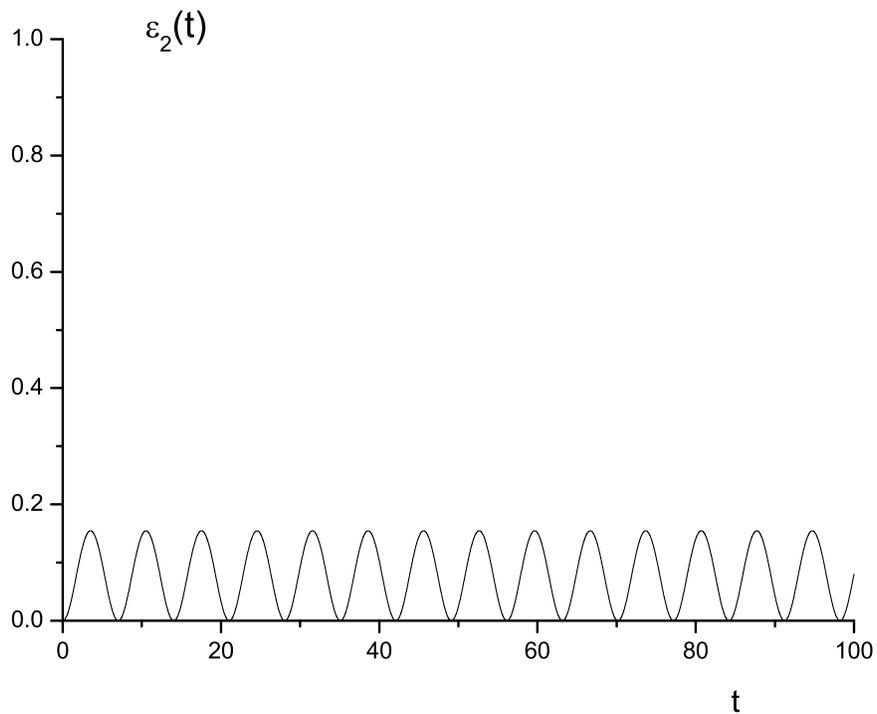,height=5in}}
\caption{ Evolutional entanglement production, quantified by the measure 
$\ep_2(t)$, in the mode-locked regime, with $b=0.3$. In this and in all 
following figures, time is measured in units of $\al^{-1}$.}
\label{fig:Fig.1} 
\end{figure}

\newpage

\begin{figure}[ht]
\centerline{\psfig{file=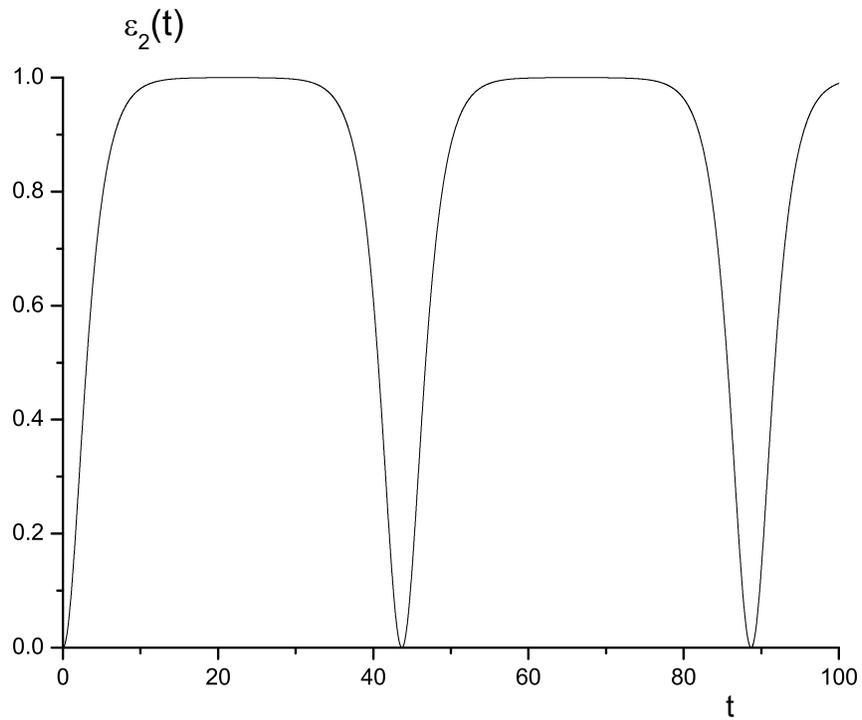,height=5in}}
\caption{The measure $\ep_2(t)$ for the boundary between the
mode-locked and mode-unlocked regimes, when $b=b_c=0.497764$.}
\label{fig:Fig.2}
\end{figure}

\newpage

\begin{figure}[ht]
\centerline{\psfig{file=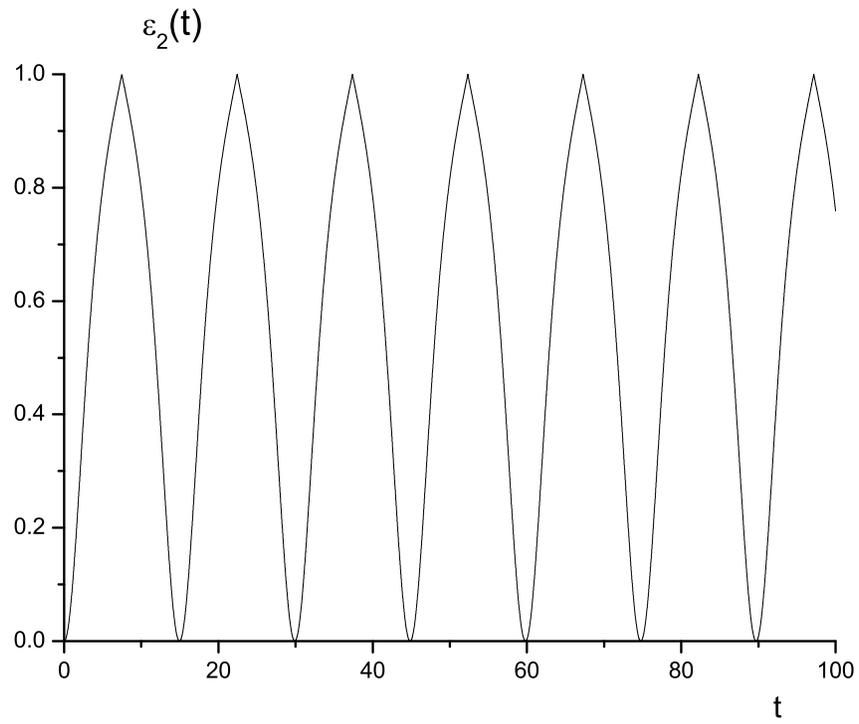,height=5in}}
\caption{Entanglement production in the mod-unlocked regime, with
$b=0.5$.}
\label{fig:Fig.3}
\end{figure}

\newpage

\begin{figure}[ht]
\centerline{\psfig{file=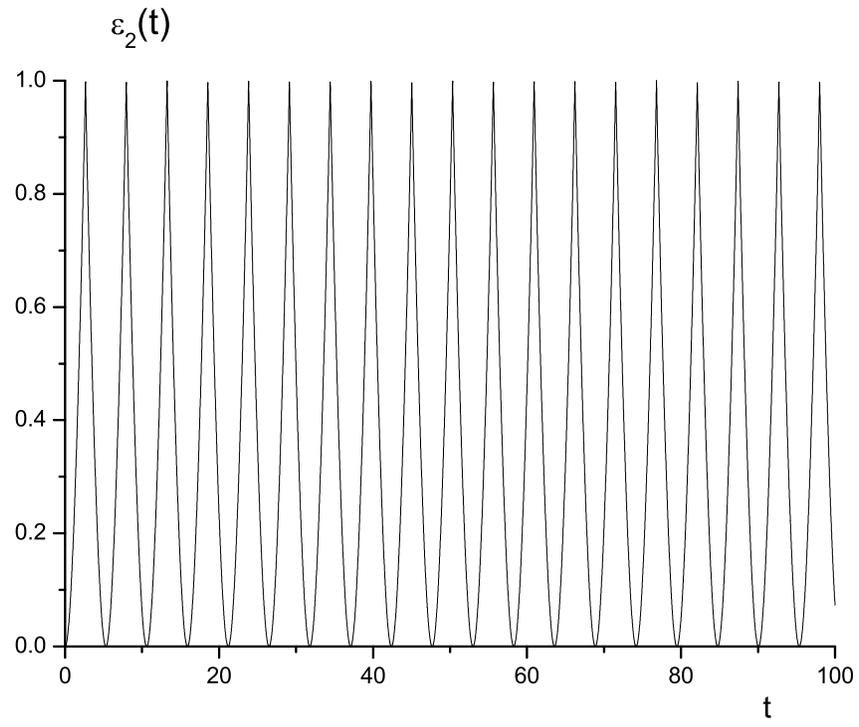,height=5in}}
\caption{Drastic shortenning of the period of $\ep_2(t)$ for $b=0.7$.}
\label{fig:Fig.4}
\end{figure}

\newpage

\begin{figure}[ht]
\centerline{\psfig{file=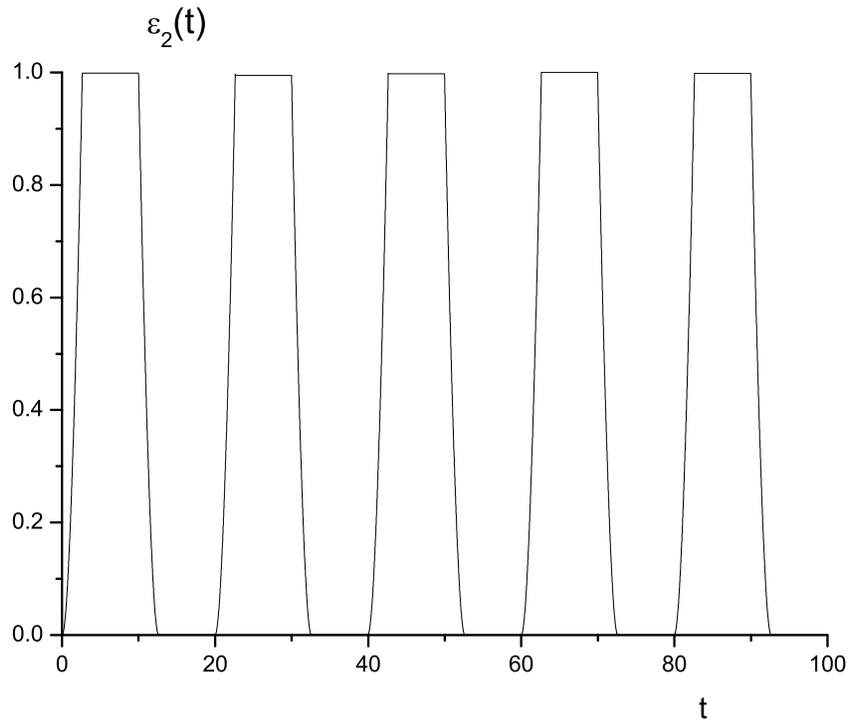,height=5in}}
\caption{Regulated equidistant pulses of $\ep_2(t)$, formed by switching 
on and off the resonant field, with $b=0.7$, so that $\ep_2(t)$ equals one 
during the time intervals $\Dlt t=7.35$ (in units of $\al^{-1}$), and it 
equals zero during the same intervals $\Dlt t=7.35$.}
\label{fig:Fig.5}
\end{figure}

\newpage

\begin{figure}[ht]
\centerline{\psfig{file=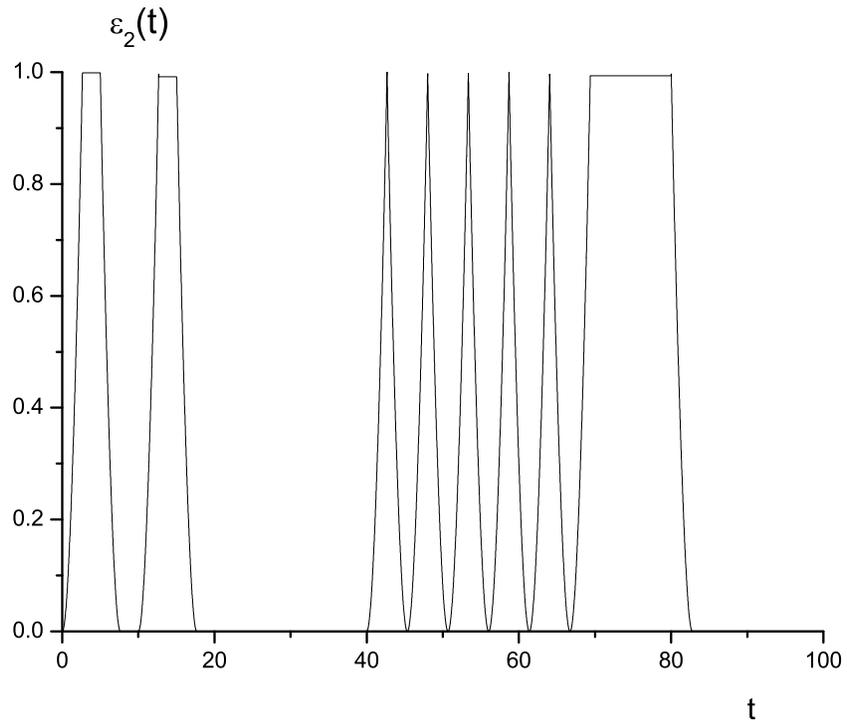,height=5in}}
\caption{Nonequidistant pulses of $\ep_2(t)$, created by switching on 
and off the pumping field, with $b=0.7$, at nonequal time intervals.}
\label{fig:Fig.6}
\end{figure}

\newpage

\begin{figure}[ht]
\centerline{\psfig{file=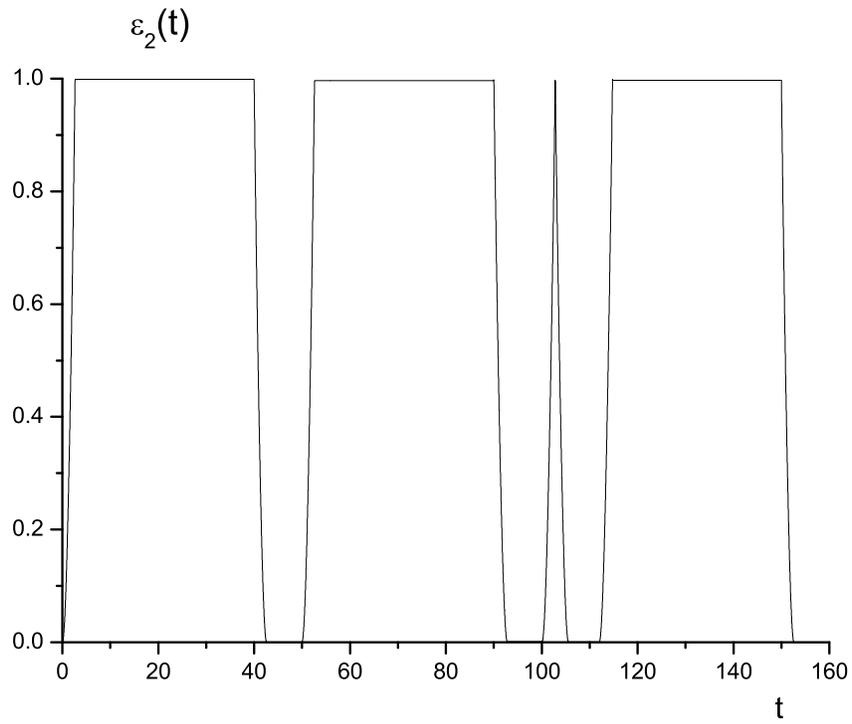,height=5in}}
\caption{Regulated pulses of $\ep_2(t)$, for the same $b=0.7$, as in 
Fig. 6, but for essentially different moments of switching on and off
the pumping field.}
\label{fig:Fig.7}
\end{figure}

\end{document}